\documentclass[aip,floats,superscriptaddress]{revtex4}

\usepackage{hyperref}
\usepackage{graphicx}
\usepackage{epstopdf}
\usepackage{amssymb}
\usepackage{amsmath,bm}
\usepackage{psfrag}
\usepackage{epsfig}
\usepackage{float}
\usepackage{bm}

\begin{document}

\title{A mesoscopic model of nanoclusters self-assembly on a graphene Moir\'e}

\author{Mikhail Khenner\footnote{Corresponding
author. E-mail: mikhail.khenner@wku.edu.}}
\affiliation{Department of Mathematics, Western Kentucky University, Bowling Green, KY 42101, USA}
\affiliation{Applied Physics Institute, Western Kentucky University, Bowling Green, KY 42101, USA}
\author{Lars Hebenstiel}
\affiliation{Department of Physics, Western Kentucky University, Bowling Green, KY 42101, USA}
\affiliation{Department of Mathematics, Western Kentucky University, Bowling Green, KY 42101, USA}

\begin{abstract}
\noindent

A continuum, post-deposition mesoscopic model of a Moir\'e-regulated self-assembly of metal nanoclusters on a twisted bilayer graphene is presented. 
Quasi-two-dimensional nanocluster-like steady states at a low adsorbate coverage are analytically determined for Pt, Ni, and Pb adsorbates, 
pointing that nanoclusters self-assemble at the Moir\'e cells centers. This is followed by the computations of nanoclusters self-assembly dynamics. 
Differences in the self-assembly efficiency for three chosen metals are highlighted across three typical values of an initial 
submonolayer coverage and for three temperature regimes. Accounting for the adsorption potential of metal atoms onto graphene leads to 
a significantly faster nanoclusters self-assembly and has a transient impact on the nanoclusters morphologies. A model extensions to the cases of
nanoclusters self-assembly on a Moir\'e formed by a monolayer graphene over a metal substrate, and the electromigration-guided self-assembly 
on such Moir\'e are proposed.  

\end{abstract}

\date{\today}
\maketitle


\section{Introduction}
\label{Intro}

Since the pioneer 2006 publication by N'Diaye \textit{et al.} \cite{DBFM}, experimental studies of a Moir\'e-regulated self-assembly of metal nanoclusters on 
graphene had become a prolific research area. Several comprehensive recent reviews of the progress made in the past decade were published 
\cite{TringidesReview,KumarReview,RuffinoReview}. Nanoclusters self-assembly on graphene was attempted for many metals, such as 
Ir \cite{DBFM,JRPPG}, Pt \cite{DGBMCM,PGHLG,ZGG,DJ}, W \cite{DGBMCM}, Re \cite{DGBMCM}, Ru \cite{EHBRLWHE,FRC}, Ni \cite{SBZRDF}, Pd \cite{ZGG,FKYWEHL}, Co \cite{ZGG}, Rh \cite{ZGG,STHSX}, 
Au \cite{ZGG,STHSX}, Pb \cite{LXYZJFDG}, and Cs \cite{PLRMBK}. Also, this method of epitaxial self-assembly is gaining popularity for organic 
substances (as reviewed in Ref. \cite{YKLKLP}) and C60 molecules \cite{LYZYBGL}. 

It is well known that a rotational misalignment of two stacked graphene layers (the arrangement that is known as twisted bilayer graphene, or TBG) results in a periodically corrugated surface known as Moir\'e 
superlattice \cite{DXS,JJB}.
When a monolayer graphene is deposited on a non-graphene single-crystal substrate, the Moir\'e effect of a rotational misalignment may be amplified by the in-plane strain 
arising from the lattice mismatch \cite{WB,MSPOM,EGW}.
A 
Moir\'e superlattice 
presents a complex potential landscape for diffusing adatoms.
Besides, the adsorption on 
graphene is highly selective \cite{TringidesReview,PGHLG,SBZRDF,STHSX}.
Consequently, nanoclusters self-assembly on a graphene Moir\'e is a complicated non-equilibrium process that 
depends on a plethora of thermodynamic and kinetic factors \cite{TringidesReview}. 

Despite significant advances in experiment,  
modeling and computation of nanoclusters self-assembly on graphene to-date is limited, to our knowledge, to a single Kinetic Monte-Carlo (KMC) study \cite{EHBRLWHE} 
and a single molecular dynamics study \cite{FRC}. 
The more broad KMC study \cite{EHBRLWHE} models the random deposition of isolated Ru atoms and their diffusion that occurs via hopping on the monolayer graphene/Ru surface. 
The rate of an adatom hopping between neighboring adsorption sites $i$ and $f$ is selected to have
the Arrhenius form $h_{i\rightarrow f}\sim \exp{\left(-E_{i\rightarrow f}/k T\right)}$, where $E_{i\rightarrow f}$ is the activation barrier and $k T$ the Boltzmann's factor.
The form of the activation barrier is chosen so that the 
density of adatoms would be maximized in the regions that have the strongest adatom-graphene binding
within a Moir\'e cell. 
Then the local cluster nucleation rate at the cite $f$ is given by the local hop rate rate 
$h_{i\rightarrow f}$ times the square of the local adatom density. 
Computations using this detailed atomistic 
model provided values for nanoclusters density and size that reasonably matched the experimentally observed values. Also, consistent with experiment 
at $T=309$ K and with deposition fluxes ranging from $0.01$ to $0.1$ ML/min,
at most one Ru nanocluster occupies each Moir\'e cell, and nanoclusters form only in the fcc region of a cell.

In this paper we propose a continuum model for computational studies of nanoclusters self-assembly on a graphene Moir\'e. 
Our model corresponds to a post-deposition regime, where the diffusive transport between Moir\'e cells is significant.
The model is rooted in the established mesoscopic continuum model framework for the adsorbate diffusion \cite{VK,LBVK,CV,CTT,W}. 
In the mesoscopic framework, the inter-atomic interactions are included via terms in a governing differential equation(s).
Although understandably not as detailed as
the atomistic studies, this model operates on a diffusion time scale and allows for uninterrupted tracking of the process of nanoclusters self-assembly 
from the random initial coverage to dense clusters. 
These computations are low cost, with the typical execution time on a desktop computer around 1 hr, and thus they allow to efficiently probe the 
differences in self-assembly for various combinations of adsorbates and Moir\'es.
As the example of the model application, we compute the self-assembly of Pt, Ni and Pb nanoclusters
on a fairly simple Moir\'e of a 
TBG, as a function of the annealing temperature, initial coverage, applied strain, and the 
strength of the Moir\'e potential.

\section{The Model}
\label{Model}


For the adatoms undergoing diffusion on graphene in a Moir\'e energy landscape $\epsilon_m(x,y)$, the dimensionless aerial density, or coverage 
$\rho(x,y,t)$ evolves according to the mass conservation \cite{CTT,W,SuoLu,LuKim,JW}:
\begin{equation}
\frac{\partial \rho}{\partial t} =   - \Omega \bm{\nabla} \cdot \bm{J} =  
-\Omega\bm{\nabla} \cdot \left[-\frac{\nu \mathcal{D} M}{k T}\rho\left(1-\rho\right)\bm{\nabla}\left\{\mu(\rho)+\delta \epsilon_m-U(\rho)\right\}\right],
\quad 0<\rho < 1\; \mbox{monolayer (ML)}, \label{rho-eq}
\end{equation}
where $\Omega$ is the area of the atom cross-section, $\nu$ the constant aerial density of adsorption sites, 
$\bm{J}$ the surface diffusion flux, 
$\mathcal{D}$ the diffusion coefficient, $M$ the diffusional mobility (assumed constant), 
$\mu(\rho)$ the adatoms chemical potential, $0<\epsilon_m(x,y)<1$ the dimensionless Moir\'e potential, $\delta$ the
magnitude of the Moir\'e potential with the dimensions of energy, $U(\rho)$ the adsorbate-adsorbate interaction potential, and
$\bm{\nabla}=(\partial_x, \partial_y)$. The factor $1-\rho$ in $\bm{J}$ accounts for the fact that the diffusion flow can pass only through the vacant lattice 
sites (a finite occupancy effect) \cite{VK,LBVK,CV,Painter,CTT,W}.  
The chemical potential $\mu(\rho)$, as always, is given by thermodynamics:
\begin{equation}
\mu(\rho) = \Omega \frac{\partial \gamma}{\partial \rho},
\label{mu}
\end{equation}
where
\begin{equation}
\gamma=k T \nu \left[\left(1-\rho\right)\ln \left(1-\rho\right) + \rho\ln \rho\right]
\label{gamma}
\end{equation}
is the free energy of an adsorbate layer, or the surface energy density, expressed by the ideal solution model \cite{CTT,W,VK}. 
For the rest of the paper we suppress the unit of coverage, writing simply, say,  $\rho_0=0.01$ instead of $\rho_0=0.01$ ML.
 
Eq. (\ref{rho-eq}) further requires the adsorbate-adsorbate interaction potential $U(\rho)$ and the Moir\'e potential $\epsilon_m(x,y)$.
These potentials are immediately introduced in Sections \ref{UModel} and \ref{MoireModel}, and the derivation of an adsorbate diffusion equation is finalized
in sections \ref{DiffEq} and \ref{DiffEqenh}. 

The starting Eq. (\ref{rho-eq}) is a nonlinear, fourth-order 
partial differential equation of a forced Cahn-Hilliard (CH) type.  The force is given by the gradient of the total potential, 
$\bm{F}=\bm{\nabla}\left\{\delta \epsilon_m-U(\rho)\right\}$. Spatially forced CH equations were used for modeling self-assembly in solid binary monolayers on 
elastic substrates \cite{SuoLu,LuKim} and 
a phase separation in binary fluids \cite{KK}. 
In this paper, since phase separation is not relevant to the process of self-assembly of a single-component metal adsorbate, the free energy (\ref{gamma}) 
is expressed by the ideal solution model, instead of the regular solution model \cite{SuoLu,LuKim,JW}. We also remark that it is quite common to include 
a ``coefficient" function of the dependent variable, such as $\rho\left(1-\rho\right)$ in Eq. (\ref{rho-eq}), in the mobility $M$. 
Derivation of this function via coarse-graining a 
microscopic balance of diffusion fluxes is transparently presented in Ref. \cite{Painter}.

The specialization of the potentials for modeling metal adsorbate diffusion and self-assembly  
on a supported graphene is one aspect that makes our mesoscopic formulation distinctly different from the published mesoscopic models \cite{VK,LBVK,CV,CTT,W}. 
These early models are for an abstract adsorbate and substrate, and they involve only one primitive (idealized) potential (either a substrate potential, 
or an adsorbate-adsorbate potential). Also, the focus of our modeling is the self-assembly process \emph{per se}, rather than the derivation 
(via nonequilibrium statistical mechanics and coarse-graining) of a mesoscopic governing equations from various microscopic dynamics. 
Note that Eq. (\ref{rho-eq}) is formulated using only the macroscopic concepts of the diffusion flux, the chemical potential, etc. as was done, for instance, 
in Refs. \cite{CTT,W}. We do not interrogate a question whether a mesoscopic dynamics is the proper limit of a "true" microscopic one since, 
for instance, in Ref. \cite{CTT} both derivations, e.g. using either the macroscopic concepts, or the coarse-graining from a microscopic level, are attempted
and it is shown that they lead to the same governing evolution equation for the adsorbate coverage. To our knowledge, only Ref. \cite{AV} is similar 
in its focus to this paper, as in that article a mesoscopic equation is derived and used to model nanoscale self-assembly dynamics of a submonolayer 
film.\footnote{Though it must be noted that the model of Ref. \cite{AV} only very loosely corresponds to a bimetallic system, Pb/Cu(111), that is claims to model, 
since the model parameters and even the form of a potential are not optimized, i.e. they are generic.}

\subsection{Adsorbate-adsorbate interaction potential}
\label{UModel}


The general form of the adsorbate-adsorbate interaction potential $U$ is given by
\begin{equation}
U=-\int_a^\infty u(r)\rho\left(\bm{r}_1\right) dr,
\label{Upot}
\end{equation}
where $u(r)$ is the pair interaction potential of two adsorbate atoms separated by the (dimensionless) distance $r=|\bm{r}_1-\bm{r}_2|$  \cite{VK,CTT}.
$\bm{r}_1$ and $\bm{r}_2$ are the time-dependent dimensionless position vectors of the atoms.
If the fixed graphene lattice spacing $a_g$ is chosen for the distance unit, then the lower integration limit in Eq. (\ref{Upot}), $a=a_{ads}/a_g$ 
(where $a_{ads}$ is the lattice spacing of the adsorbate). Thus $a$ also is dimensionless, whereas the pair interaction potential $u(r)$, and thus the  
potential $U$, have the dimension of energy.
Expanding $\rho\left(\bm{r}_1\right)$ in the Taylor series about 
$\bm{r}_2$ yields:
\begin{equation}
U=-\int_a^\infty u(r)\left[\sum_{m=0}^\infty \frac{r^m}{m!}\bm{\nabla}^m \rho\left(\bm{r}_2\right)\right] dr=
-\int_a^\infty u(r)\left[\rho\left(\bm{r}_2\right)+ r \bm{\nabla}\rho\left(\bm{r}_2\right)+\frac{r^2}{2}\bm{\nabla}^2\rho\left(\bm{r}_2\right)+...\right] dr.
\end{equation}
Assuming that $u(r)$ is spherically symmetric 
gives
\begin{equation}
U\approx-\rho\int_a^\infty u(r) dr - \frac{\bm{\nabla}^2\rho}{2}\int_a^\infty  r^2 u(r) dr. 
\label{Uform}
\end{equation}

We chose the Sutton-Chen (SC) model \cite{SuttonChen} for the pair interaction potential: 
\begin{equation}
u(r)=\epsilon\left[\left(\frac{a}{r}\right)^n-2c\left(\frac{a}{r}\right)^{m/2}\right].
\label{u-of-r}
\end{equation}
Here $\epsilon$ is the potential magnitude with the dimension of energy, and $c$, $m$ and $n$ are dimensionless.
At $n>m$, $u(r)$ is attractive at long distances and repulsive at very short distances, preventing the (unphysical) collision of the atoms, 
which are thought of as hard spheres.
Provided the adsorbate and graphene lattice spacings, in effect, provided value of $a$,
this model may be fitted to represent adatoms interaction within a particular adsorbate by optimizing the values of four parameters: 
$\epsilon$, $c$ (a positive real number), and the positive integers $m,\ n$. The optimization was done by Sutton \& Chen \cite{SuttonChen}
for a number of metals. For Pt, Ni, and Pb, in Table \ref{T1} we list values of $\epsilon$, $a$, $c$, $m$ and $n$ from Ref. \cite{SuttonChen}.  
As is clear from the form of Eq. (\ref{u-of-r}), the SC potential belongs to a family of interatomic Lennard-Jones potentials \cite{RM,LC,HVFN}.
Our choice of the SC potential is guided by the fact that this potential was optimized 
with high accuracy for major elemental metals using the same process, 
which gives us the confidence and allows the comparison of certain self-assembly characteristics for the three chosen metals.

Substituting Eq. (\ref{u-of-r}) into Eq. (\ref{Uform}) yields
\begin{equation}
U = -A\rho-B\bm{\nabla}^2 \rho,
\label{U}
\end{equation}
where 
\begin{equation}
A=\epsilon a\left(\frac{1}{n-1}+\frac{4c}{2-m}\right),\quad
B=\frac{\epsilon}{2} a^3 \left(\frac{1}{n-3}+\frac{4c}{6-m}\right)
\label{ABCD}
\end{equation}
are the zero and second moments of the potential. Due to their proportionality to $\epsilon$, these moments have the dimension of energy. 
Note that values of $A$ and $B$ are calculated using values of $\epsilon$, $a$, $c$, $m$ and $n$ from Table \ref{T1}, which results in $A$ and $B$ negative.

\subsection{Moir\'e potential}
\label{MoireModel}

To construct a realistic Moir\'e potential $\epsilon_m(x,y)$, first a Moir\'e lattice needs to be constructed. Constructions of Moir\'e lattices of various complexities 
are described in several articles \cite{H,SMKB,MPSO,CWCF,WMCF}. Here, we adopt the construction in Cosma \textit{et al.} \cite{CWCF} and Wallbank \textit{et al.} \cite{WMCF}, 
since that construction (i) is in the real space, rather than in the reciprocal space, and (ii) it incorporates the strain in the explicit and transparent 
fashion. 

According to Cosma \textit{et al.} \cite{CWCF} and Wallbank \textit{et al.} \cite{WMCF}, a Moir\'e lattice that is formed upon a deposition of a single 
graphene layer on a substrate is magnified from the atomic scale of the substrate 
lattice constant by the large factor of the order of $\left(\delta'^2+\theta^2-w'^2\right)^{-1/2}$, where $\delta'\ll 1$ is related to the lattice mismatch, 
$\theta$ is the rotation angle of the graphene lattice with respect to the substrate lattice, and $w'$ is related to the 
uniaxial strain in the substrate or the graphene. Precisely, the Moir\'e lattice vectors $\mathbf{A}_k$
are magnified from the graphene lattice vectors $\mathbf{a}_k,\; k=0,1,...,5$ by the matrix $\hat M$, as follows:
\begin{equation}
\bm{A}_k=\hat M \bm{a}_k,\quad \hat M= \frac{1}{\delta'^2+\theta^2-w'^2} 
\begin{pmatrix}
\delta'+\left(\ell_y^2-\ell_x^2\right)w' & \theta-2\ell_x\ell_y w'\\
-\theta-2\ell_x\ell_y w' & \delta'-\left(\ell_y^2-\ell_x^2\right)w'
\end{pmatrix},
\label{Am}
\end{equation}
where $\bm{\ell}=\left(\ell_x,\ell_y\right)$ is the principal axis of the strain tensor, $\delta'=\delta+w(1-\sigma)/2$, $w'=w(1+\sigma)/2$, and
$\sigma$ is the Poisson ratio. Here $\delta$ and $w$ are the lattice mismatch and uniaxial strain, respectively. The form of Eq. (\ref{Am}) assumes that the strain is applied to the substrate. 
The set of six dimensionless vectors $\mathbf{a}_k$ is obtained by a $k\pi/3$ rotation of $\bm{a}_0=(1,0)$. 
Without loss of generality, we chose $\ell_x=1,\ \ell_y=0$. In this work, we assume that the substrate is graphene $(\delta=0)$, 
thus we consider only the Moir\'e of TBG.
The Poisson ratio of graphene is 0.165.  

\begin{figure}[H]
\vspace{-0.2cm}
\centering
\includegraphics[width=4.5in]{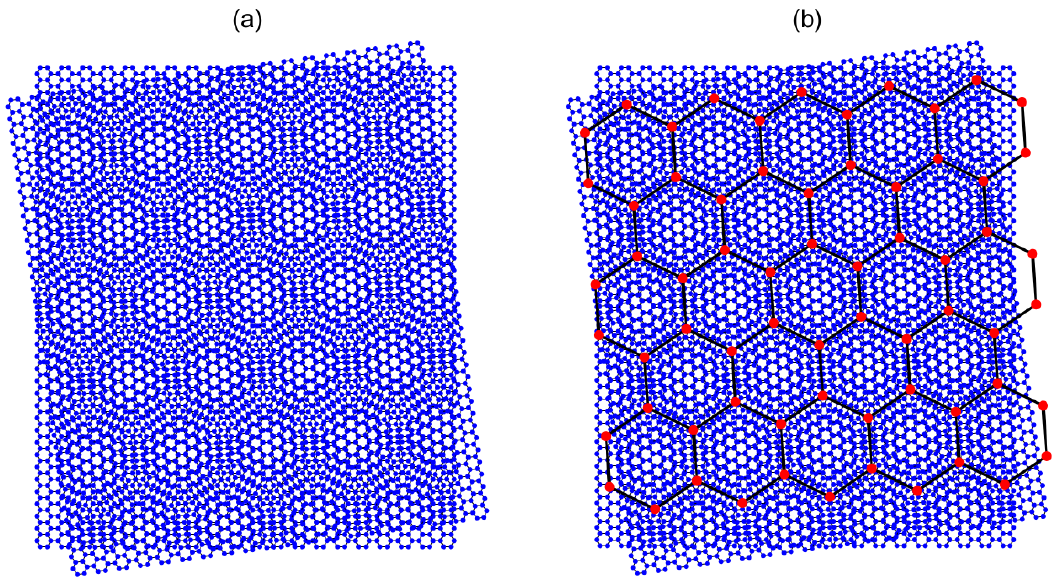}
\vspace{-0.15cm}
\caption{(a) Moir\'e pattern on TBG. Counter-clockwise rotation angle of the top graphene layer is $\theta = 8^\circ$. Blue dots represent carbon atoms 
(the size is not to scale). 
Length of a bond between two neighbor carbon atoms is equal to one, since $a_g$ is used as the length scale.
(b) Corresponding Moir\'e lattice constructed using Eq. (\ref{Am}). Lattice vertices are marked by large red dots for visibility only.
}
\label{Fig1}
\end{figure}

Fig. \ref{Fig1}(a) shows the Moir\'e pattern that results from the $8^\circ$ counter-clockwise rotation of the unstrained graphene lattice (top) 
with respect to the unstrained graphene substrate (bottom), whereas in Fig. \ref{Fig1}(b) the Moir\'e lattice constructed using Eq. (\ref{Am}) with 
values $w=0$ and $\theta=8^\circ$ is overlapped over Fig. \ref{Fig1}(a). A perfect match can be observed, which validates
Eq. (\ref{Am}). Since there is no comparison in Refs. \cite{CWCF} and \cite{WMCF} of Eq. (\ref{Am}) to a directly constructed Moir\'e pattern 
such as Fig. \ref{Fig1}(a), we considered such validation necessary before proceeding to the construction of the Moir\'e potential. 

\begin{figure}[H]
\vspace{-0.2cm}
\centering
\includegraphics[width=6.5in]{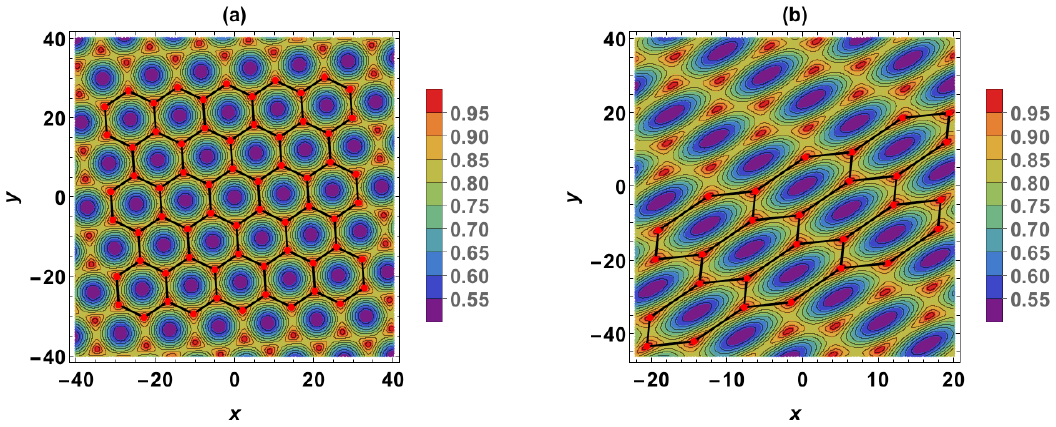}
\vspace{-0.15cm}
\caption{(a) Moir\'e potential, Eqs. (\ref{Mpotent1}), (\ref{Mpotent2}) ($D_m=0.5$) with the Moir\'e lattice from Fig. \ref{Fig1}(b). $w=0$.
This potential is referred to as $\epsilon^A_m(x,y)$ in the remainder of the paper.  
(b) Moir\'e potential ($D_m=0.5$) and the Moir\'e lattice at $w=0.1$.  This potential is referred to as $\epsilon^B_m(x,y)$ in the remainder of the paper. 
}
\label{Fig2}
\end{figure}

The Moir\'e potential is given by:
\begin{eqnarray}
\epsilon_m(x,y)&=&1-\frac{D_m}{3}\sqrt{3+2\left[\cos{\bm{x}\cdot \bm{c}_1}+\cos{\bm{x}\cdot \bm{c}_2}+\cos{\bm{x}\cdot \left(\bm{c}_1-\bm{c}_2\right)}\right]}, \label{Mpotent1}\\
\bm{c}_1&=&\hat M^{-1}\left(1/2,\sqrt{3}/2\right)^T,\quad \bm{c}_2=\hat M^{-1}\left(-1/2,\sqrt{3}/2\right)^T,\quad \bm{x}=(-\frac{4\pi}{3}x,\frac{4\pi}{3}y). 
\label{Mpotent2}
\end{eqnarray}
Here $0<D_m<1$ is the depth of the potential well and the matrix $\hat M$ is to be computed with the same values of the parameters that are used to construct 
the Moir\'e lattice. The potential form (\ref{Mpotent1}) is motivated by the energy spectrum of a honeycomb lattice in the tight-binding approximation \cite{honey}. In Fig. \ref{Fig2}(a) the constructed lattice from Fig. \ref{Fig1}(b) is overlapped 
with the matching potential; in Fig. \ref{Fig2}(b) the Moir\'e lattice and the potential are re-computed assuming applied 10\% strain ($w=0.1$) 
to the substrate graphene. Perfect match again is observed in both cases. The potentials in Figures \ref{Fig2}(a,b) are labeled $\epsilon^A_m(x,y)$ and
$\epsilon^B_m(x,y)$, respectively, and are used in Eq. (\ref{rho-eq-final}) to model nanoclusters self-assembly on the corresponding Moir\'e.  


\subsection{Basic adsorbate diffusion equation}
\label{DiffEq}

We are now in the position, where we can finish the formulation of the basic diffusion equation that models adsorbate self-assembly on a graphene Moir\'e.

Combining Eqs. (\ref{rho-eq}) - (\ref{gamma}) and (\ref{U}), and choosing graphene lattice spacing $a_g$ for the length unit, 
$a_g^2/\Omega^2 \nu^2 \mathcal{D} M$ for the time unit, and $k T$ for the energy unit, we obtain: 
\begin{eqnarray}
\frac{\partial \rho}{\partial t} = -\bm{\nabla}\cdot \bm{J} &=& -\bm{\nabla} \cdot \left[-\rho\left(1-\rho\right)\bm{\nabla}\left( \ln \frac{\rho}{1-\rho}+f_1\rho+
f_2\bm{\nabla}^2\rho+f_3\epsilon_m \right) \right] \nonumber \\ 
&=& \bm{\nabla}^2 \rho + \bm{\nabla}\cdot \left[\rho\left(1-\rho\right)\left(f_1\bm{\nabla}\rho+
f_2\bm{\nabla}\bm{\nabla}^2\rho+f_3\bm{\nabla} \epsilon_m\right)\right].
\label{rho-eq-final}
\end{eqnarray}
Although $\epsilon_m(x,y)$ is described by the explicit formula (\ref{Mpotent2}), here we do not substitute that formula in Eq. (\ref{rho-eq-final}), 
as this results in a complicated expression. Eqs.  (\ref{Mpotent2}) and (\ref{rho-eq-final}) form the closed dimensionless equation 
that describes the evolution of the adsorbate coverage $\rho(x,y,t)$. Notice that $\epsilon_m(x,y)$ and its first and second partial derivatives are 
continuous, as required by the operator $\bm{\nabla}\cdot \bm{\nabla} \epsilon_m\equiv \bm{\nabla}^2 \epsilon_m$.
  
The dimensionless parameters in Eq. (\ref{rho-eq-final}) are $f_1=AN/k T$, $f_2=BN/k T$, $f_3=\delta N/k T$, where 
$A<0$ and $B<0$ are the first and the second moments of the adsorbate-adsorbate interaction potential, as introduced in Eqs. (\ref{ABCD}).
Thus $f_1,\ f_2 < 0,\ f_3>0$. 
$N=1/\Omega \nu=4/\pi d^2 \nu$ is dimensionless, where $d$ is the atomic diameter 
(the area of an atom cross-section $\Omega=\pi r^2=\pi d^2/4$, where $r$ is the atomic radius).
$N$ is the fourth independent dimensionless parameter 
that we chose to absorb in the other three dimensionless parameters $f_1$, $f_2$, and $f_3$.
For calculation of $N$, we define 
$\nu=S^{-1}$, where $S=(3\sqrt{3}/2)a_g^2$  is the area of a (hexagonal) graphene cell formed by six adjacent carbon atoms. 
This assumes one adsorption site per graphene cell. Here $a_g=2.46\times 10^{-8}$cm.  
Fixed $\delta=2.4\times 10^{-12}$erg is chosen for the calculation of $f_3$. (For reference, $\delta/k T=58.508,\ 29.918,\ 19.637$ at $T=20^\circ,\ 300^\circ$ and  $600^\circ$, 
respectively.) We list values of $d$, $f_1$, $f_2$, and $f_3$ in Table \ref{T1} for Pt, Ni, and Pb at room temperature.  
 
Eq. (\ref{rho-eq-final}) describes a Fickian diffusion of the adsorbate (first term on the right-hand side), mediated by the adsorbate-adsorbate interaction 
(first and second terms in the bracket) and by the substrate potential (third term in the bracket). It is appropriate to remark here that accounting in 
the derivation for the second moment, $B$, of the interaction potential is necessary, 
since the corresponding $f_2$-term 
in Eq. (\ref{rho-eq-final}) (and in Eq. (\ref{rho-eq-final_enh}) formulated below) provides a short-wavelength cutoff, which is necessary for the linear stability with 
respect to a short-wavelength perturbations of any constant base state. In other words, without the $f_2$-term the evolution of the coverage is unstable 
($\rho\rightarrow \infty$) due to amplification of a small short-wavelength perturbations, which typically are of thermal nature and thus are present in any adsorbate/substrate system.
Thus $f_2$ fulfills the role of the dimensionless CH gradient energy coefficient.

\begin{table}[!ht]
\centering

\begin{tabular}
{|c|c|c|c|c|c|c|c|c|c|c|}
\hline
				 
			\rule[-2mm]{0mm}{6mm} Adsorbate & $\epsilon$, erg & $a$ & $c$ & $m$ & $n$ & $d$, cm & $f_1$ & $f_2$ & $f_3$ &  $f_1/f_2$ \\
			\hline
                        \hline
			\rule[-2mm]{0mm}{6mm} Pt & $3.172\times 10^{-14}$ & 1.593 & 34.408  & 8 & 10 & $3.54\times 10^{-8}$ & -44.93 & -171.62 &  93.46 & 0.262\\
			\hline
                        \rule[-2mm]{0mm}{6mm} Ni & $2.55\times 10^{-14}$ & 1.431 & 39.426  & 7 & 9 & $2.98\times 10^{-8}$ & -63.0 & -323.38  & 131.89 & 0.195  \\
			\hline
				\rule[-2mm]{0mm}{6mm} Pb & $9.02\times 10^{-15}$ & 2.000 &  45.77  & 7 & 10 & $3.08\times 10^{-8}$ & -16.05 & -160.91  & 123.47 & 0.1 \\
			\hline
					
\end{tabular}
\caption[\quad ...]{Dimensional ($\epsilon$) and dimensionless ($a$, $c$, $m$, $n$) parameters of the Sutton-Chen potential, the atomic diameter $d$, 
and the derived dimensionless parameters 
$f_1$, $f_2$, $f_3$ for Eqs. (\ref{rho-eq-final}) and (\ref{rho-eq-final_enh}). 
$f_1$, $f_2$, $f_3$ are shown at $T=20^\circ$. See the text for the complete descriptions of all parameters.
}
\label{T1}
\end{table}

\subsection{Enhanced adsorbate diffusion equation}
\label{DiffEqenh}

In deriving Eq. (\ref{rho-eq-final}), the surface diffusivity $\mathcal{D} M$ was assumed a constant. On a microscopic level, the atom hop rate from an occupied site to a neighbor 
vacant site is a variable that depends on the atom's local environment and the temperature. Recognizing this fact and relaxing the constant diffusivity 
assumption, whereby assuming that the coverage is not varying extremely fast, the procedure of coarse graining from a microscopic level to a mesoscopic level 
produces the following form of the dimensionless surface diffusivity \cite{VK}: 
$\exp{\left(-f_4 v(x,y)+ U\right)} \approx \exp{\left(-f_4 v(x,y)+ f_1 \rho\right)}$, where $v(x,y)$ is a spatially 
varying adsorption, or binding, potential and $f_4$ is the dimensionless parameter quantifying its strength. We can measure $f_4$ in units of $f_1$, 
thus writing $f_4=\sigma f_1$.  

Enhanced adsorbate diffusion equation now reads:
\begin{equation}
\frac{\partial \rho}{\partial t} = -\bm{\nabla}\cdot \bm{J} = -\bm{\nabla} \cdot \left[-e^{|f_1|\left(\sigma v - \rho\right)}\rho\left(1-\rho\right)\bm{\nabla}\left( \ln \frac{\rho}{1-\rho}+f_1\rho+
f_2\bm{\nabla}^2\rho+f_3\epsilon_m \right) \right].
\label{rho-eq-final_enh}
\end{equation}
Eq. (\ref{rho-eq-final_enh}) formally reverts to Eq. (\ref{rho-eq-final}) when the  
exponential spatio-temporal diffusivity is replaced by one. From now on we refer to these cases as a ``variable diffusivity" and a ``constant diffusivity", 
respectively. In Sec. \ref{B}, we compare the features of nanoclusters self-assembly for these cases. 


\section{Quasi-2D steady-states at low coverage}
\label{Steady}

In this section we ask two questions: Is there a final state of the evolution of an \emph{arbitrary} initial coverage ? 
And, if there is such state (called a steady-state), what is its form, that is, what is the final $\rho(x,y)$ ? 
For evolution governed by Eq. (\ref{rho-eq-final}), we are able to answer these questions analytically (and therefore exactly), rather than numerically - 
which is a quite rare situation. The kinetics (dynamics) of the approach to a steady-state is addressed numerically in Sec. \ref{CompositionEvolve}. 

We start by assuming the solution
of Eq. (\ref{rho-eq-final}) with $\partial \rho/\partial t=0$ (since the coverage has finished to evolve, reaching a sought-after steady-state) 
and $\bm{\nabla} \rightarrow \partial/\partial x$ in the form of a power series in a small parameter $\zeta$:
\begin{equation}
\rho(x,y) = \zeta \rho_1(x,y) + \zeta^2 \rho_2(x,y) + \zeta^3 \rho_3(x,y)...,\quad \zeta\ll 1
\label{series}
\end{equation}
The form (\ref{series}) is substituted in that equation (where the zero at the left-hand side 
is represented as $0\zeta+0\zeta^2+0\zeta^3+...$), the equation is expanded, and the coefficients of the terms that have the same power of $\zeta$ on the left side and on the right 
side are equated. This procedure formally results in the infinite set of a coupled ordinary differential equations (ODEs) in the variable $x$ for 
$\rho_i,\ i=1,2,3,...$.  Limiting the derivation to $\rho_1$ and $\rho_2$, the ODEs for these quantities read:
\begin{eqnarray}
\frac{\partial^2 \rho_1}{\partial x^2}+f_3\frac{\partial}{\partial x}\left[\rho_1 \frac{\partial \epsilon_m}{\partial x}\right]&=&0,\label{st1}\\
\frac{\partial^2 \rho_2}{\partial x^2}+\frac{\partial}{\partial x}\left[f_3\left(\rho_2-\rho_1^2\right) \frac{\partial \epsilon_m}{\partial x} + 
f_1 \rho_1 \frac{\partial \rho_1}{\partial x} +f_2 \rho_1 \frac{\partial^3 \rho_1}{\partial x^3}\right]&=&0.\label{st2}
\end{eqnarray}
Assuming even symmetry, $\rho_{1,2}(-x,y)=\rho_{1,2}(x,y)$, $\epsilon_m(-x,y)=\epsilon_m(x,y)$, linear Eqs. (\ref{st1}) and (\ref{st2}) are easily integrated. 
(Recall that as far as the integration in $x$ is concerned, $y$ is the dummy variable.) The solutions read:
\begin{eqnarray}
\rho_1&=&C_1\exp{\left(-f_3 \epsilon_m\right)},\quad C_1>0\label{r1sol}\\
\rho_2&=&C_2\exp{\left(-f_3 \epsilon_m\right)}+
\frac{2-f_1}{2}\rho_1^2-f_2\left[\rho_1 \frac{\partial^2 \rho_1}{\partial x^2}-\frac{1}{2}\left(\frac{\partial \rho_1}{\partial x}\right)^2\right]. \label{r2sol1}
\end{eqnarray}
We may take $C_2=C_1$, then
\begin{equation}
\rho_2=\rho_1+\frac{2-f_1}{2}\rho_1^2-f_2\left[\rho_1 \frac{\partial^2 \rho_1}{\partial x^2}-\frac{1}{2}\left(\frac{\partial \rho_1}{\partial x}\right)^2\right]. \label{r2sol}
\end{equation}
Thus the quasi-2D steady-state solution of Eq. (\ref{rho-eq-final}) is approximated as
\begin{equation}
\rho(x,y)=\zeta \rho_1(x,y) + \zeta^2 \rho_2(x,y),
\label{finalsteady}
\end{equation}
where $\rho_1(x,y)$ and $\rho_2(x,y)$ are given by Eqs. (\ref{r1sol}), (\ref{r2sol}). Note that, if instead of initially replacing
$\bm{\nabla}$ by $\partial/\partial x$ in Eq. (\ref{rho-eq-final}) and integrating in $x$, one replaces $\bm{\nabla}$ by $\partial/\partial y$ 
and integrates in $y$, then, Eq. (\ref{r1sol}) is unchanged, and in Eq. (\ref{r2sol}), $\partial \rho_1/\partial x \rightarrow \partial \rho_1/\partial y$ and  
$\partial^2 \rho_1/\partial x^2 \rightarrow \partial^2 \rho_1/\partial^2 y$. 

\begin{figure}[H]
\vspace{-0.2cm}
\centering
\includegraphics[width=6.5in]{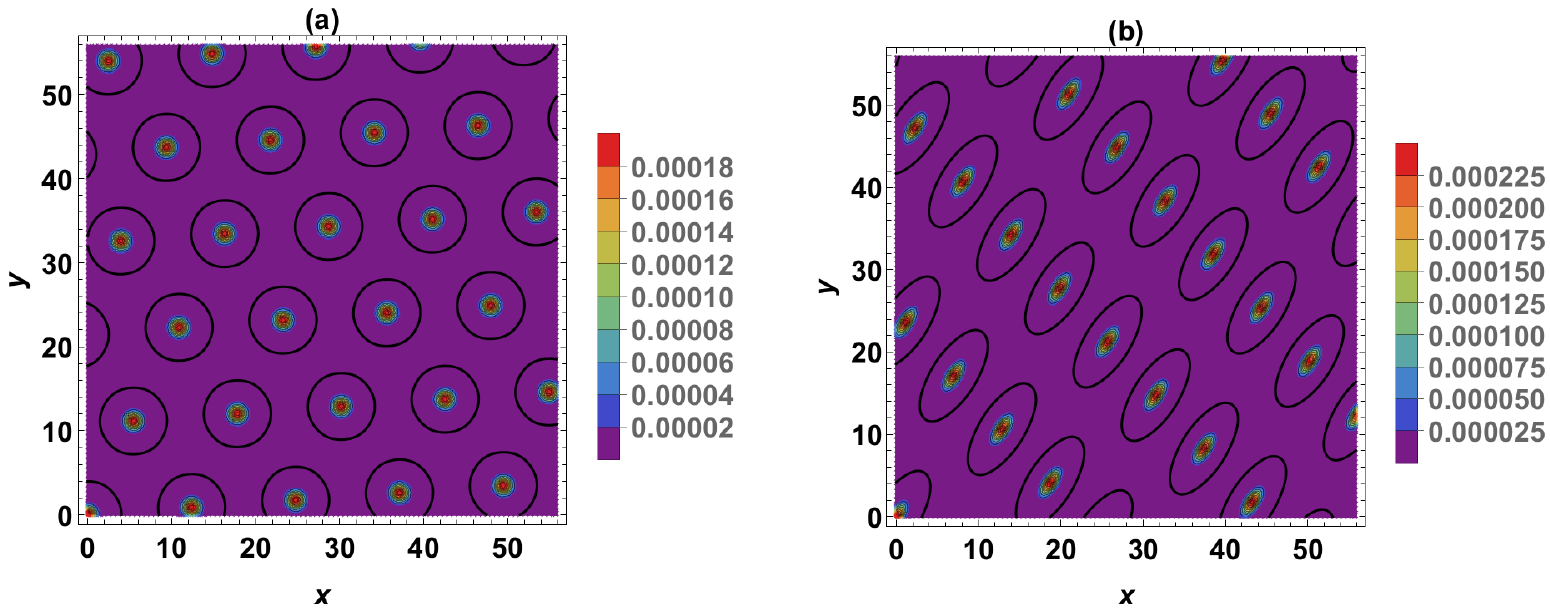}
\vspace{-0.15cm}
\caption{Quasi-2D steady-state coverage, Eq. (\ref{finalsteady}), for Pt adsorbate at $T=20^\circ$ and the Moir\'e potentials (a) $\epsilon^A_m(x,y)$ and (b) $\epsilon^B_m(x,y)$. 
Black closed contours mark the 0.7 level set of the potentials. At the centers of these contours, the potentials are at the minimum value 0.5. $\zeta=0.001,\ C_1=4\times 10^{19}$.
}
\label{Fig2point5}
\end{figure}

Fig. \ref{Fig2point5} shows the obtained quasi-2D steady-state solution computed for Pt and the Moir\'e potentials 
$\epsilon^A_m(x,y)$ and $\epsilon^B_m(x,y)$. For Ni and Pb (plots not shown) there is no change, except, at $C_1$ fixed to $4\times 10^{19}$ as in 
Fig. \ref{Fig2point5}, a different interval of $\rho$ values on the color bars. That is, the visual appearance of the plots is unchanged. 
($C_1=4\times 10^{19}$ is arbitrarily chosen because $C_1$ value cannot be found in this 
analytic steady-state determination. Thus here, the \emph{magnitude} of $\rho$ can be adjusted to any desired value by choosing a $C_1$ value. However, 
the \emph{shape} of the function $\rho(x,y)$ is uniquely determined by Eq. (\ref{finalsteady}).) It will be seen in Sec. \ref{CompositionEvolve} that the coverage indeed evolves 
to the nanocluster states
aligned with the minimas of the Moir\'e potential, that closely resemble those in Fig. \ref{Fig2point5}
(see Figures \ref{Fig3} and \ref{Fig4}).

\section{Computational results}
\label{CompositionEvolve}

\subsection{Long-time evolution of the coverage toward the formation of a self-assembled nanoclusters}

In this section we reveal the dynamics of an adsorbate coverage evolution over a long time interval. For this purpose, it is sufficient to compute 
using Eq. (\ref{rho-eq-final}), since the dynamics of the basic Eq. (\ref{rho-eq-final}) and enhanced Eq. (\ref{rho-eq-final_enh}) 
differ only on shorter time scales, as will be described in Sec. \ref{B}.

Computations of Eq. (\ref{rho-eq-final}) are performed on the fixed, square, bi-periodic domain using the Method of Lines framework with a pseudospectral 
spatial discretization.   
Various random initial conditions for the coverage $\rho$ are chosen, with the mean at $\rho_0=0.01,\ 0.1$, or $0.5$ (very low, low, and high) and the maximum 
deviation of $15\%$ from the mean value. Conservation of mass, $ \iint_R \rho \,dx\,dy$, was checked during coverage evolution. 

\begin{figure}[H]
\vspace{-0.2cm}
\centering
\includegraphics[width=6.5in]{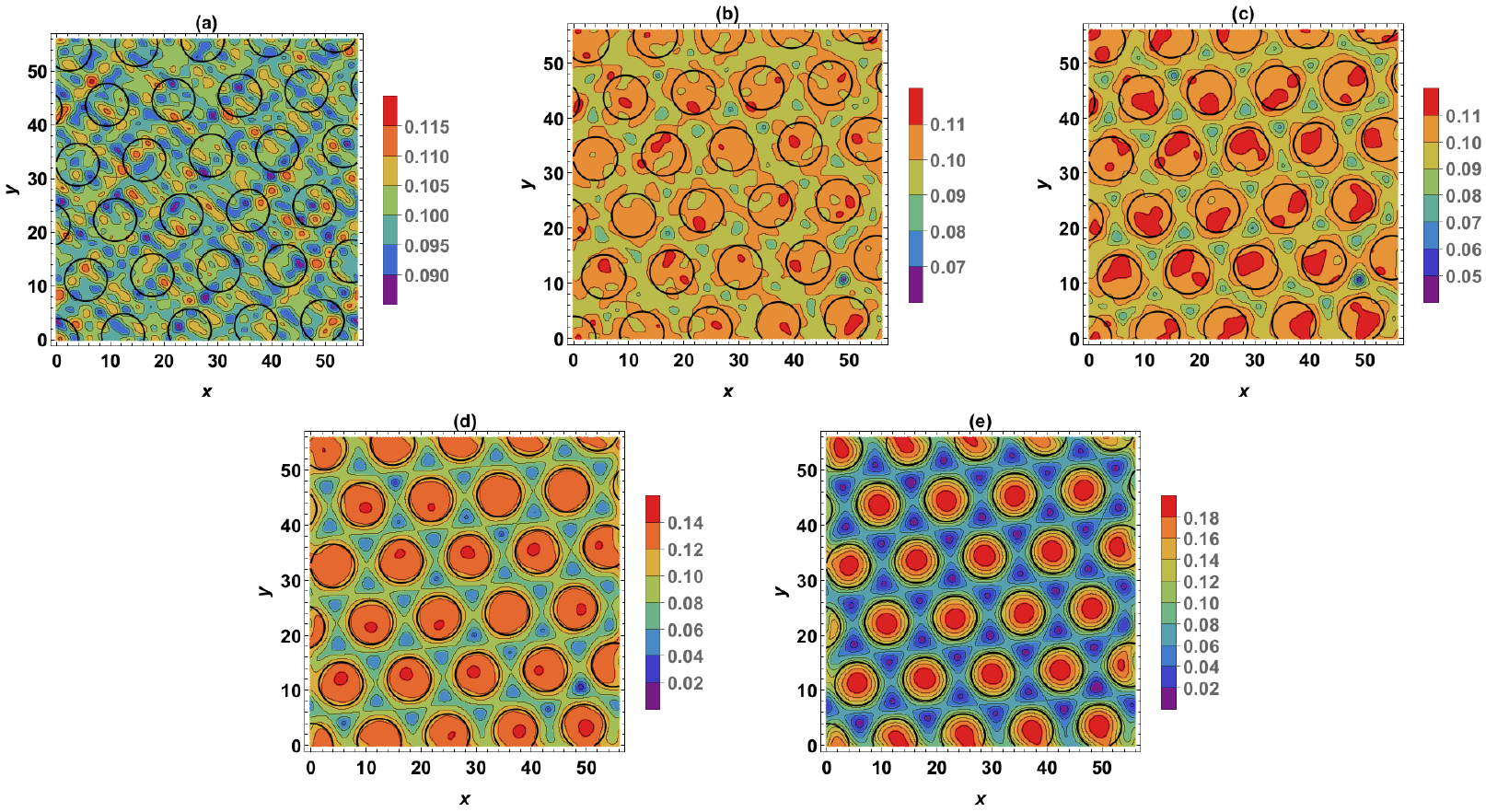}
\vspace{-0.15cm}
\caption{Coverage $\rho$ of Pt adsorbate at (a): $t=0$ (the initial condition), (b): $t=0.01$, (c): $t=0.02$, (d): $t=0.07$ and (e): $t=0.14$. 
Black closed contours mark the 0.7 level set of $\epsilon^A_m(x,y)$. $\rho_0=0.1,\ T=20^\circ$.
}
\label{Fig3}
\end{figure}

\begin{figure}[H]
\vspace{-0.2cm}
\centering
\includegraphics[width=6.5in]{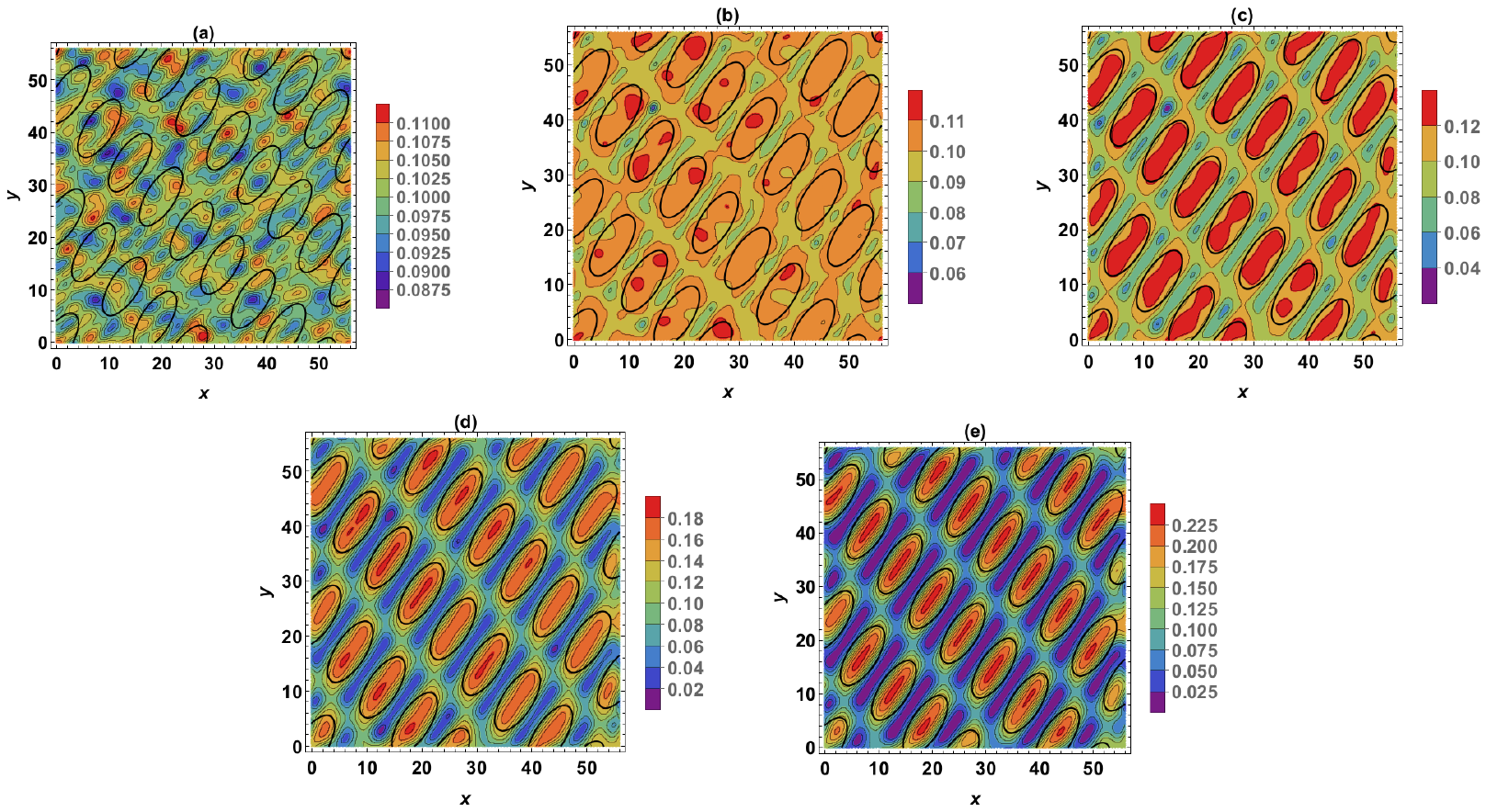}
\vspace{-0.15cm}
\caption{Coverage $\rho$ of Pt adsorbate at (a): $t=0$ (the initial condition), (b): $t=0.01$, (c): $t=0.04$, (d): $t=0.1$ and (e): $t=0.16$. 
Black closed contours mark the 0.7 level set of $\epsilon^B_m(x,y)$. $\rho_0=0.1,\ T=20^\circ$.
}
\label{Fig4}
\end{figure}

Figures \ref{Fig3} and \ref{Fig4} show the snapshots of Pt coverage at the increasing times, for evolution in Moir\'e potentials $\epsilon^A_m(x,y)$ 
and $\epsilon^B_m(x,y)$, respectively. These computations were executed at $\rho_0=0.1$ and $T=20^\circ$.
We observe the progression of the adatoms assembly at the spatial locations of the Moir\'e potential minima, which ultimately leaves bare other areas of 
the top graphene layer. Due to the continuum nature of the model, the computation has to be stopped once the coverage reaches zero locally; this is the final time, $t_f$.
The pathway to the final self-assembled state at $t=t_f$ as depicted in Figures \ref{Fig3}(e) and \ref{Fig4}(e) is visually similar for other initial coverages 
and temperatures, and for another two adsorbates, e.g. Ni and Pb at various coverages and temperatures. However, the maximum coverage at 
the Moir\'e cell center at $t=t_f$ is not the same, and we are quantifying it below.


\begin{figure}[H]
\vspace{-0.2cm}
\centering
\includegraphics[width=6.5in]{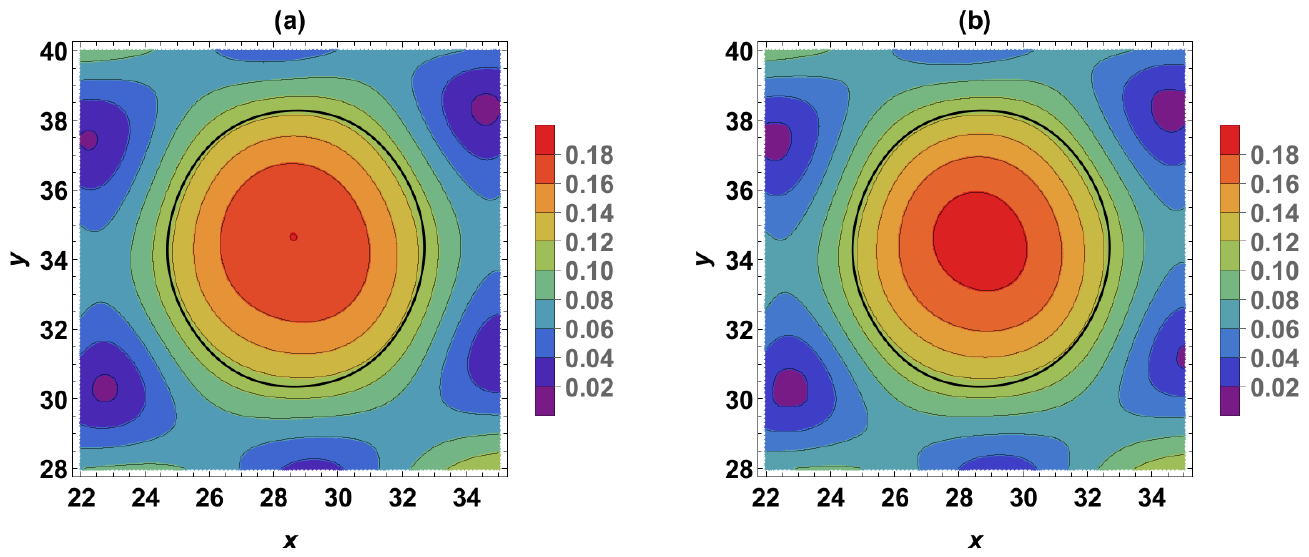}
\vspace{-0.15cm}
\caption{Coverage $\rho$ of Pt adsorbate at (a): $t=0.12$ and (b): $t=0.13$. 
Black closed contours mark the 0.7 level set of $\epsilon^A_m(x,y)$. $\rho_0=0.1,\ T=20^\circ$.
}
\label{Fig5}
\end{figure}

Fig. \ref{Fig5} shows that the shape of the coverage distribution within the Moir\'e cell conforms to the shape of the cell.
The largest $\rho$ level spreads out from the cell center, and after the area enclosed by the largest $\rho$ reaches approximately 1/5 of the cell area,
a new, larger  "nucleus" appears at the center (Fig. \ref{Fig5}(a)). This nucleus in turn grows laterally and the cycle repeats.   

\begin{figure}[H]
\vspace{-0.2cm}
\centering
\includegraphics[width=6.5in]{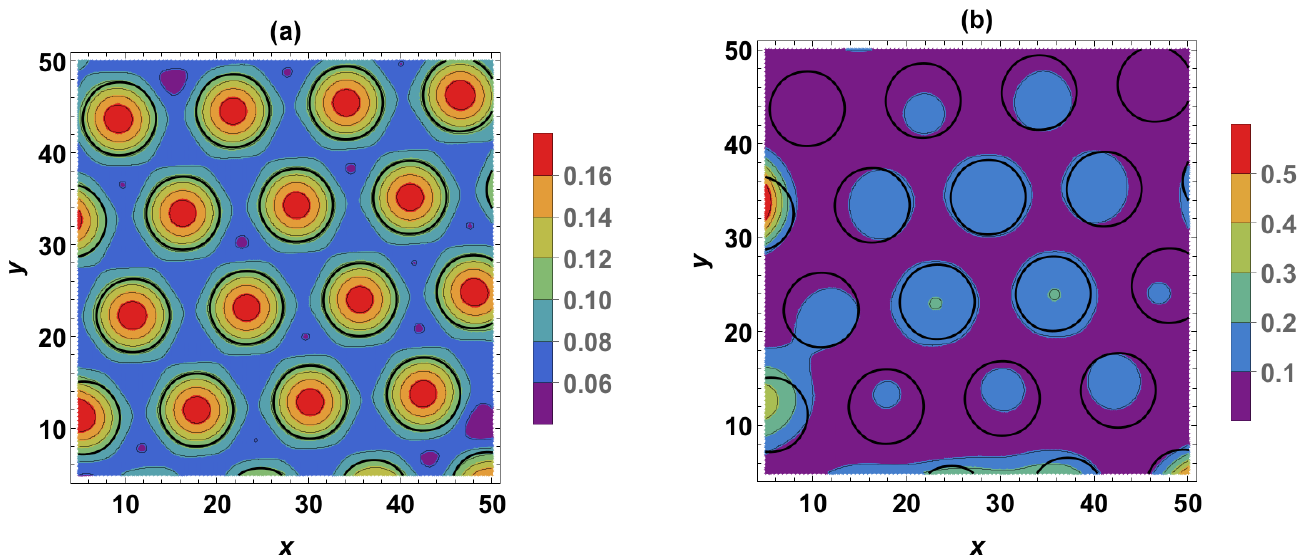}
\vspace{-0.15cm}
\caption{Coverage $\rho$ of Pt adsorbate at (a): $t=4$ and (b): $t=40$. 
Black closed contours mark the 0.985 level set of $\epsilon^A_m(x,y)$. $\rho_0=0.1,\ T=20^\circ$.
}
\label{Fig6}
\end{figure}

Next, we computed nanoclusters self-assembly (at $T=20^\circ$ only) in the Moir\'e potentials $\epsilon^{A,B}_m(x,y)$ of a decreasing depth $D$ in order to 
pinpoint the smallest $D$ at which a self-assembly still occurs. These computations show that for three chosen adsorbates, a self-assembly is 
effective at $D$ larger than 0.03. However, as expected, longer times are required at smaller $D$ values. In Fig. \ref{Fig6} the results for Pt at 
$D=0.025$ are shown. It can be seen that for a while after the initiation of diffusion the adatoms do assemble at the Moir\'e cells, however 
this process is not sustained in the long run.
Eventually, the already assembled nanoclusters disassemble and in the center of the computational domain the coverage reverts to the 
equilibrium value $\rho=0.1$.

\begin{figure}[H]
\vspace{-0.2cm}
\centering
\includegraphics[width=3.5in]{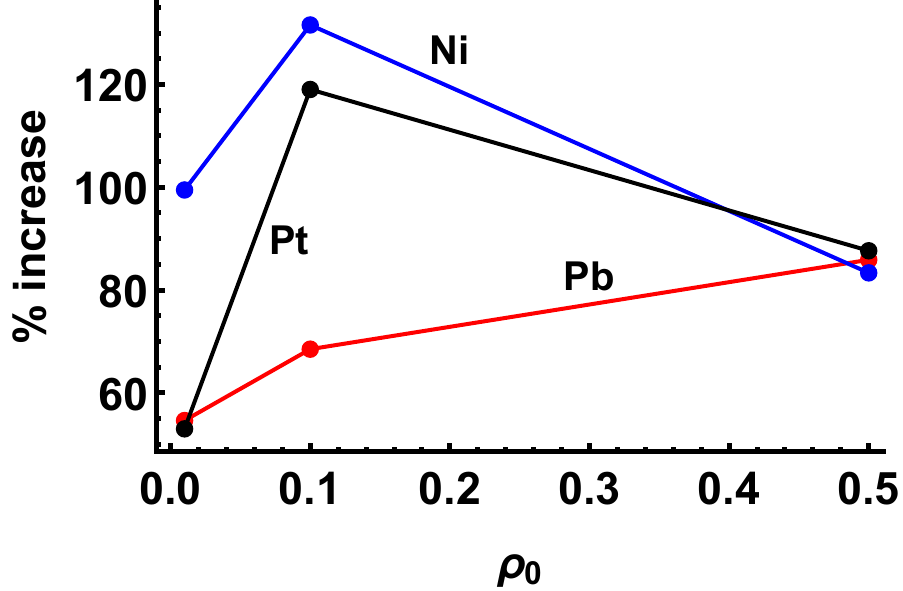}
\vspace{-0.15cm}
\caption{Efficiency of nanoclusters self-assembly vs. the three initial coverage levels, $\rho_0=0.01, 0.1$, and 0.5.  $T=20^\circ$. 
}
\label{Fig7}
\end{figure}

Fig. \ref{Fig7} compares the efficiency of nanoclusters self-assembly for three adsorbates at $T=20^\circ$. Efficiency is defined as the maximum \%-increase
of the coverage in the Moir\'e cell center relative to the initial coverage. Ni nanoclusters assembly most efficiently, 
while Pb nanoclusters assemble the least efficiently. The largest relative increase of the coverage occurs at the low initial 
coverage $\rho_0=0.1$. At the high initial coverage the difference between the three adsorbates vanishes. The largest overall efficiency is 131\%
(for Ni at $\rho_0=0.1$), which means that the highest coverage in the Moir\'e cell center is 0.231. The close look at these results suggests that 
the highest efficiency is achieved for a metal adsorbate that has the largest magnitude of $f_1$, and therefore the largest magnitude of the first moment, $A$,  
of the adsorbate-adsorbate interaction potential. Larger efficiency for Ni also follows from the comparison of the cohesive energies.
Using Eqs. (6) and (7) in Ref. \cite{SuttonChen}, we obtained $E_c\approx \frac{(2n-m)m^2}{2n^2m}c^2\epsilon$, which for Ni gives the largest value, 
$E_c^{Ni}=1.88\times 10^{-11}$ erg/atom, followed by $E_c^{Pt}=1.8\times 10^{-11}$ erg/atom and $E_c^{Pb}=8.6\times 10^{-12}$ erg/atom.

\begin{figure}[H]
\vspace{-0.2cm}
\centering
\includegraphics[width=6.5in]{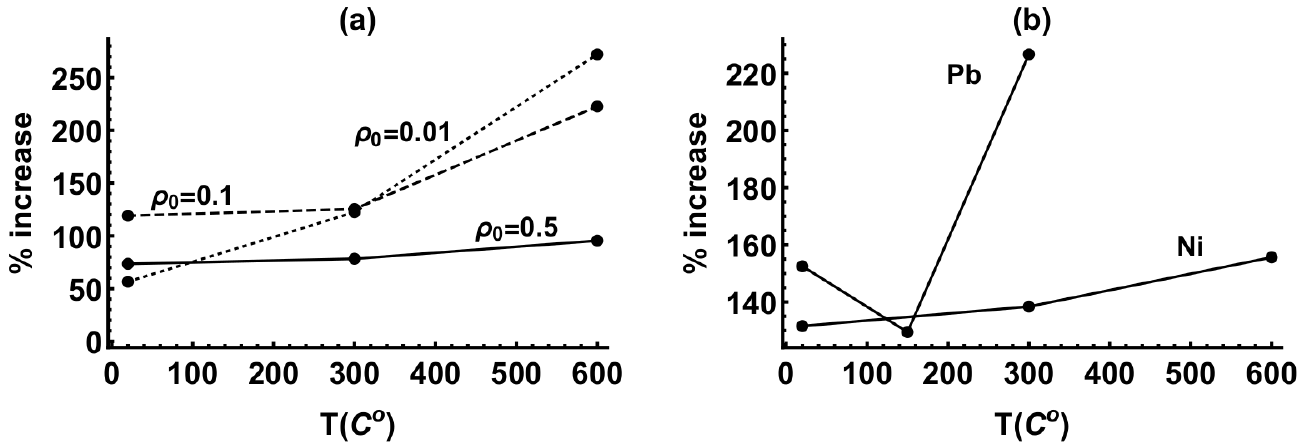}
\vspace{-0.15cm}
\caption{(a) Efficiency of Pt nanoclusters self-assembly vs. temperature for the three initial coverage levels. 
(b) Efficiency of Ni and Pb nanoclusters self-assembly vs. temperature at $\rho_0=0.1$.
}
\label{Fig8}
\end{figure}

Fig. \ref{Fig8}(a) shows the efficiency vs. temperature for Pt, again for the three initial coverage levels. Efficiency increases as $T$ increases
,
and the steepest increase occurs at low initial coverage. At the high coverage the temperature has the minimal impact on efficiency. 
Finally, \ref{Fig8}(b) shows the efficiency vs. temperature for Ni and Pb at low coverage. For Ni, the efficiency increases as $T$ increases, 
similar to Pt. However, comparing Ni curve to Pt curve in Fig. \ref{Fig8}(a) at $\rho_0=0.1$ (the dashed curve), we see that at high temperature Ni assembles 
less efficiently than Pt. For Pb, which has much lower melting temperature of 327 C$^\circ$, the efficiency dips around 150 C$^\circ$, then increases. 
Similar trends for Ni and Pb are expected at very low coverage and at high coverage.
Notice that the dimensionless parameters
$f_1,\ f_2$ and $f_3$ in  Eq. (\ref{rho-eq-final}) all are inversely proportional to $T$, thus for all three adsorbates the increased efficiency at 
higher $T$  is attributed to 
the overall higher mobility of the adsorbate, e.g. the relative increase of the Fickian term $\bm{\nabla}^2\rho$.

\subsection{Short-time effects of the adsorption potential on nanoclusters self-assembly}
\label{B}

In this section we attempt to characterize the effect of the adsorption potential on self-assembly. To this end, we adopt Eq. (\ref{rho-eq-final_enh})
for computations and choose a special initial condition for the coverage $\rho$.

\begin{figure}[H]
\vspace{-0.2cm}
\centering
\includegraphics[width=6.5in]{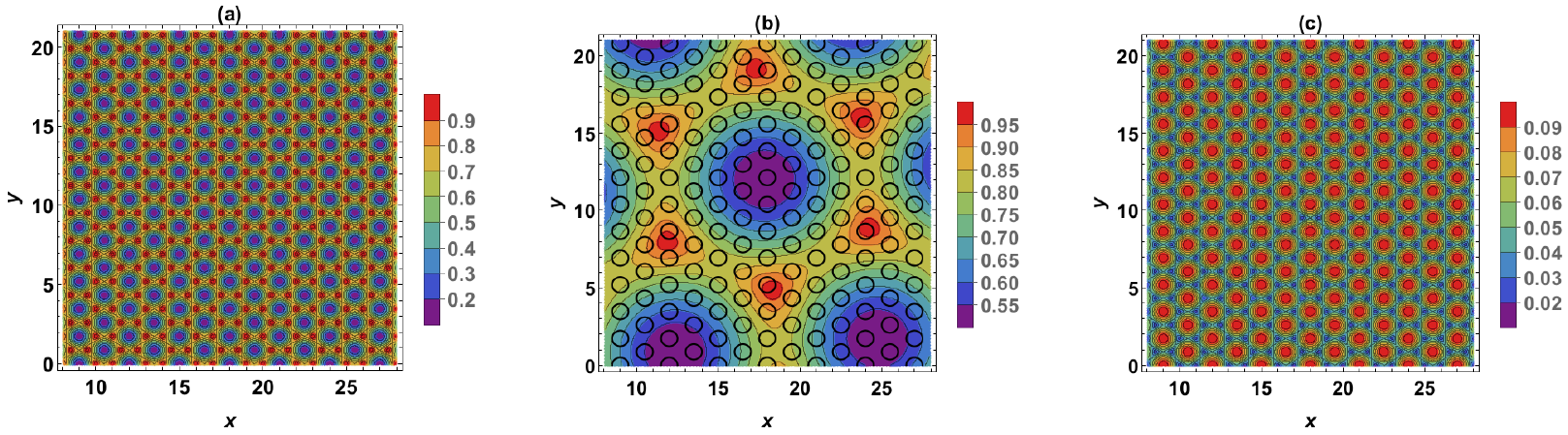}
\vspace{-0.15cm}
\caption{(a) The adsorption potential $v(x,y)$ with $D_a=0.5$. (b) Overlap of the Moir\'e potential $\epsilon_m^{(A)}$ with $D_m=0.5$ 
and the adsorption potential $v$ with $D_a=0.5$. Black closed contours mark the 0.7 level set of $v$.
(c) Initial coverage of Pt adsorbate.
}
\label{Fig9}
\end{figure}

We take the adsorption potential $v$ in Eq. (\ref{rho-eq-final_enh}) in the form (\ref{Mpotent1}), (\ref{Mpotent2}), where the matrix $\hat M$ is replaced by the identity matrix and the 
potential well depth $D_m$, by $D_a$. Thus $v$ spatially varies on the scale of a single graphene cell. As shown in Fig. \ref{Fig9}(a), the minimas of $v$ 
are attained at graphene's hollow sites, that is, at the centers of the hexagonal graphene cells. Fig. \ref{Fig9}(b) overlaps $v$ and the Moir\'e potential 
$\epsilon_m^{A}$; the domain for this visualization is selected such that a single Moir\'e cell is located roughly at the domain center. 
It is seen that there is multiple minimas (and maximas) of $v$ per the single cell of the Moir\'e potential. To summarize, Eq. (\ref{rho-eq-final_enh}) 
involves two spatially varying potentials, $v$ and $\epsilon_m$, which vary on very different length scales (short and long, respectively), but the functional 
form of the variation is the same. For computations that we next describe, the initial coverage of Pt adsorbate, e.g. $\rho(x,y,0)$, maximizes 
at the graphene's hollow sites, that is, at the location of the minimas of $v$. This mimics the preferential adsorption at the hollow sites. 
In other words, apart from the amplitude, $\rho(x,y,0)$ is the inverted $v$.
This is shown in Fig.  \ref{Fig9}(c).

\begin{figure}[H]
\vspace{-0.2cm}
\centering
\includegraphics[width=6.5in]{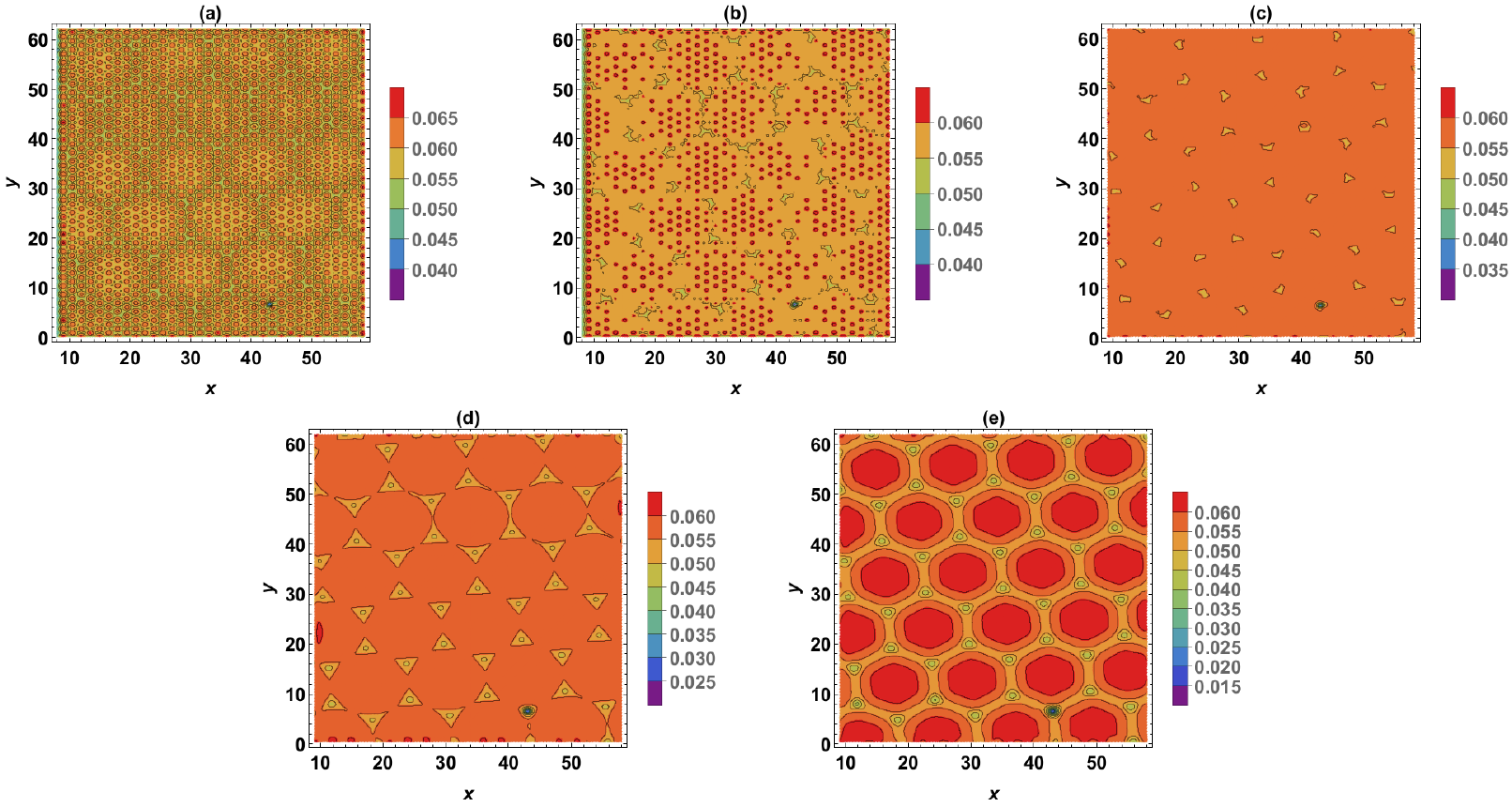}
\vspace{-0.15cm}
\caption{Coverage $\rho$ of Pt adsorbate at $T=20^\circ$, computed using Eq. (\ref{rho-eq-final}) with the Moir\'e potential $\epsilon_m^A$ and the initial condition in 
Fig. \ref{Fig9}(c), at (a): $t=0.0018$, (b): $t=0.0024$, (c): $t=0.003$, (d): $t=0.008$ and (e): $t=0.017$. 
}
\label{Fig10}
\end{figure}

It was briefly remarked above that the diffusivity in Eq. (\ref{rho-eq-final_enh}) varies in space and in time; the latter variation is due to the 
dependence of the coverage $\rho$ on $t$. However, the effect of the coverage on the diffusivity is minor compared to the effect of the 
adsorption potential, unless $\sigma$ is small. Indeed, for (spatial) variations of $v$ and $\rho$ in Fig. \ref{Fig9}(a,c), 
$\exp{\left(-|f_1| \rho\right)} < \exp{\left(\sigma |f_1| v\right)}$ at, roughly, $\sigma > 0.1$. This is the regime in which we conduct the computations.
The results discussed below were obtained with $\sigma=0.5$. (Using instead small $\sigma$ values ($\sigma\sim 0.01$), or small negative values, 
$-0.2<\sigma<0$,  resulted in the computed coverage evolution that 
is very similar to Fig. \ref{Fig10}, e.g. the one that was computed with the constant diffusivity.) 

\begin{figure}[H]
\vspace{-0.2cm}
\centering
\includegraphics[width=6.5in]{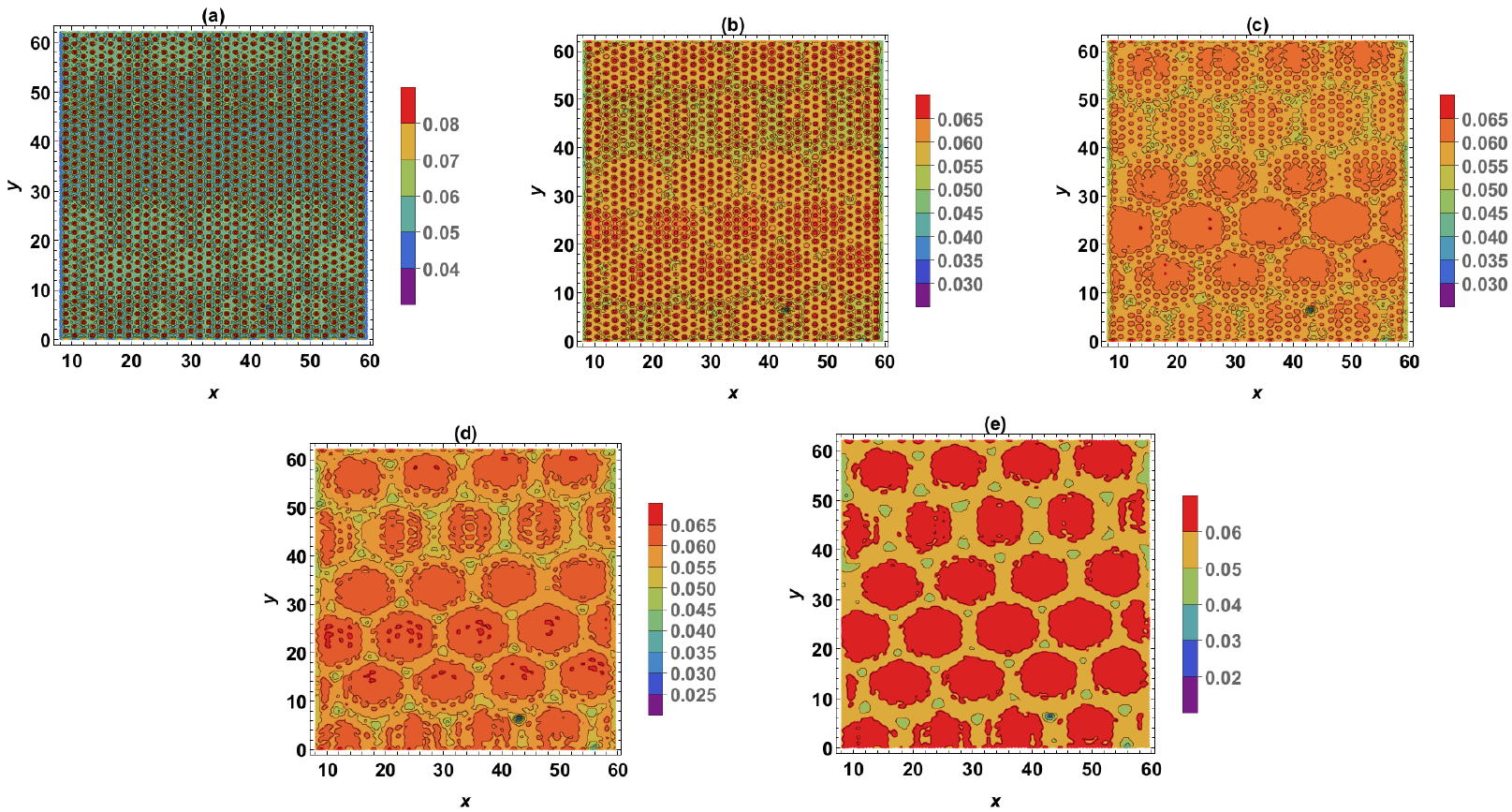}
\vspace{-0.15cm}
\caption{Coverage $\rho$ of Pt adsorbate $T=20^\circ$, computed using Eq. (\ref{rho-eq-final_enh}) with the Moir\'e potential $\epsilon_m^A$, the adsorption potential $v$ in Fig. \ref{Fig9}(a), 
and the initial condition in Fig. \ref{Fig9}(c), at (a): $t=0.00014$, (b): $t=0.0004$, (c): $t=0.0006$, (d): $t=0.001$ and (e): $t=0.0014$. 
}
\label{Fig11}
\end{figure}

Fig. \ref{Fig11} shows the results of the computation using Eq. (\ref{rho-eq-final_enh}) and the described setup. 
For close comparison, Fig. \ref{Fig10} shows the results of the computation using Eq. (\ref{rho-eq-final}). In order to not distract from observing 
the formation of nanoclusters and their morphologies, in Figures \ref{Fig10} and \ref{Fig11} we do not show 
level sets of the Moir\'e potential. 

First, by comparing Figures \ref{Fig10}(e) and \ref{Fig11}(e) we observe that when the diffusivity is variable, the evolution of the coverage 
is one order of magnitude faster. From these Figures we also notice that large clusters that ultimately formed at the spatial locations of the Moir\'e 
potential minimas have more regular shapes when the diffusivity is constant. Not quite surprising, these shapes are hexagonal, which is a consequence 
of the initial condition that has a hexagonal symmetry as well. In both cases, the clusters continue to coarsen.    

When the diffusivity is constant, the contours of the future self-assembled nanoclusters are noticeable at the early stage of the evolution (Fig. \ref{Fig10}(a)),
e.g. when the initial higher coverage at the graphene hollow sites is still clearly seen. With the passage of time the coverage preferentially smooths out, 
attaining roughly the constant value inside the Moir\'e cells (Figures \ref{Fig10}(b,c)), whereas the coverage decreases at the vertices of the Moir\'e cells, 
whose locations correspond to the maximas of the Moir\'e potential (Fig. \ref{Fig10}(c)). These domains of low coverage expand (Fig. \ref{Fig10}(d)) 
(the diffusion flow is from the Moir\'e cell boundary inward to the cell center) and slowly
the hexagonal clusters emerge (Fig. \ref{Fig10}(e)). Notice that the time interval between the stage (a) and the stage (c) is less than  
one-fourth of the time interval from the stage (c) to the stage (e). 

When the diffusivity is variable, although the outlines of the future self-assembled nanoclusters again can be seen at the early stages 
(Figures \ref{Fig11}(a,b)), the adsorption potential pins high coverage to graphene hollow sites much longer. Nanoclusters self-assemble 
by ``merging" high coverage levels at the adjacent hollow sites, as seen in Figures \ref{Fig11}(b,c). Even when a few large clusters had already formed,
some individual hollow sites can still be seen. Large clusters continue to incorporate the individual sites (Figures \ref{Fig11}(c,d)) and thereby they grow. 
Due to this  process, the cluster's boundaries have irregular shapes, although overall the shapes are approximately hexagonal; as the time goes by, 
the hexagonal shapes are becoming more abundant (Fig. \ref{Fig11}(e)). After the last hollow site has been incorporated into a cluster, the evolution
with a variable diffusivity ceases to differ from the one with a constant diffusivity; nanoclusters coarsening starts taking place, similar to Figures 
\ref{Fig3}(c-e) and Figures \ref{Fig4}(c-e). 

\section{Discussion}
\label{Summ}

We developed, analyzed and computed a mesoscopic model of an adsorbate self-assembly by diffusion in the graphene Moir\'e potential. 
For three metal adsorbates that were
studied, the model points to Ni self-assembly as the most efficient. Also, for all three metal adsorbates the self-assembly is most efficient at low
coverage, and there is a threshold strength of the Moir\'e potential below which a self-assembly is not sustained. 
Model enhancement by incorporation of an adsorption potential leads to a significantly faster nanoclusters self-assembly and 
has a transient impact on nanoclusters morphologies. 
Overall, our modeling contributes to better understanding of the physical factors that govern a Moir\'e-regulated self-assembly.  

Results of our modeling mimic certain aspects of the experiment. For instance, AFM measurements performed in Ref. \cite{ZGG} (reviewed in Ref. \cite{RuffinoReview}) 
show that on the graphene/Ru Moir\'e, Pt, Rh, Pd, Co, and Au nanoclusters self-assemble at fcc sites, which are the global energy minimas 
of the Moir\'e potential. This is also characteristic of Ru nanoclusters on the graphene/Ru Moir\'e \cite{EHBRLWHE}, as was mentioned in the Introduction. 
These conclusions also are supported by DFT calculations, see Ref. \cite{STHSX} (also reviewed in Ref. \cite{RuffinoReview}). 
Nanoclusters in our model also form at the global energy minimas of the Moir\'e potential for TBG.
One route toward better quantitative matching of the model to experiment is the formulation of a functional form of the 
Moir\'e potential that matches the experimentally observed localization of nanoclusters within a Moir\'e cell for a particular adsorbate/graphene/substrate
combination (work in progress).
Another interesting study would be of 
electromigration-driven self-assembly on graphene \cite{BRHB,SV}. In Ref. \cite{BRHB}, a reversible electromigration-driven mass transport of Au and Al 
atoms and nanoclusters along graphene is experimentally demonstrated, and in Ref. \cite{SV} a tight-binding model of electromigration of an adsorbate on graphene
is developed; it is shown that even for atoms that are strongly (covalently) bound to graphene the drift velocity can reach 1 cm/s at (moderate) 
electrical current densities of 1 A/mm and temperatures of 300-500 K. 
The interplay of a Moir\'e potential and electromigration may results in an unusual spatial localization of nanoclusters on graphene.  
One of the authors made recent contributions to studies of electromigration effects 
on nanostructuring of metal surfaces \cite{MySurfSci,MyPRMat}. 

Compared to KMC modeling, our mesoscopic approach has the important advantage of naturally enabling the tracking of the nanoclusters morphology evolution from
the early to the late stages of the nanocluster formation. This capability is either absent, or was not reported in the only KMC study \cite{EHBRLWHE} 
that we are aware of.
Moreover, that study is limited to nucleation and agglomeration of a few-atom clusters, and ripening of these small clusters into
larger clusters was not simulated. By contrast, there is no limit on the number of atoms in the cluster in our mesoscopic approach. The spatial footprint of the 
clusters computed by KMC modeling \cite{EHBRLWHE} resembles a small red dot-like clusters that cover a single Moir\'e cell taken from our 
Figures \ref{Fig10}(b) and \ref{Fig11}(b), that is, only a fraction of the full evolution time interval 
to equilibrium was accessed in the KMC modeling.  MD simulations 
\cite{FRC}  were limited to an even smaller, in fact, a tiny time interval of 5.5 ns, which only allowed to evaluate the effective diffusion constant 
by calculating the mean square displacement of the center of mass of the initial atomic configuration.     
 


\medskip
\noindent 
{\bf Data Availability.} The data that support the findings of this study are available from the corresponding author
upon reasonable request.


\begin{thebibliography}{200}

\bibitem{DBFM} A.T. N'Diaye, S. Bleikamp, P.J. Feibelman, and T. Michely, 
``Two-Dimensional Ir Cluster Lattice on a Graphene Moir\'e on Ir(111)", 
\textit{Phys. Rev. Lett.}$\;$ {\bf 97}, 215501 (2006).


\bibitem{JRPPG} M.D. Jimenez-Sanchez, C. Romero-Muniz, P. Pou, R. Perez, and J.M. Gomez-Rodriguez,
``Graphene on Rh(111): A template for growing ordered arrays of metal nanoparticles with different periodicities",
\textit{Carbon}$\;$ {\bf 173}, 1073e1081 (2021).



\bibitem{TringidesReview} X. Liu, Y. Han, J.W. Evans, A.K. Engstfeld, R.J. Behm,
M.C. Tringides, M. Hupalo, H.-Q. Lin, L. Huang, K.-M. Ho, D. Appy, P.A. Thiel, and C.-Z. Wang,  
``Growth morphology and properties of metals on graphene", 
\textit{Prog. Surf. Sci.}$\;$ {\bf 90}, 397 (2015).

\bibitem{KumarReview} A. Kumar, K. Banerjee, and P. Liljeroth,  
``Molecular assembly on two-dimensional materials", 
\textit{Nanotechnology}$\;$ {\bf 28}, 082001 (2017).

\bibitem{RuffinoReview} F. Ruffino and F. Giannazzo,  
``A Review on Metal Nanoparticles Nucleation and Growth on/in Graphene", 
\textit{Crystals}$\;$ {\bf 7}, 219 (2017).

\bibitem{DGBMCM} A.T. N'Diaye, T. Gerber, C. Busse, J. Myslivecek, J.Coraux, and T. Michely, 
``A versatile fabrication method for cluster superlattices", 
\textit{New. J. Phys.}$\;$ {\bf 11}, 103045 (2009).

\bibitem{EHBRLWHE} A.K. Engstfeld, H.E. Hoster, R.J. Behm, L.D. Roelofs, X. Liu, C.-Z. Wang, Y. Han, and J.W. Evans,
``Directed assembly of Ru nanoclusters on Ru(0001)-supported graphene: STM studies and atomistic modeling",
\textit{Phys. Rev. B}$\;$ {\bf 86}, 085442 (2012).

\bibitem{FRC} G.D. F\"{o}rster, F. Rabilloud, and F. Calvo,
``Atomistic modeling of epitaxial graphene on Ru(0001) and deposited ruthenium nanoparticles",
\textit{Phys. Rev. B}$\;$ {\bf 92}, 165425 (2015).

\bibitem{PGHLG} Y. Pan, M. Gao, L. Huang, F. Liu, and H.-J. Gao, 
``Directed self-assembly of monodispersed platinum nanoclusters on graphene Moir\'e template", 
\textit{Appl. Phys. Lett.}$\;$ {\bf 95}, 093106 (2009).

\bibitem{SBZRDF} M. Sicot, S. Bouvron, O. Zander, U. Rudiger, Vu.S. Dedkov, and M. Fonin, 
``Nucleation and growth of nickel nanoclusters on graphene Moir\'e on Rh(111)", 
\textit{Appl. Phys. Lett.}$\;$ {\bf 96}, 093115 (2010).

\bibitem{ZGG} Z. Zhou, F. Gao, and D.W. Goodman, 
``Deposition of metal clusters on single-layer graphene/Ru(0001): Factors that govern cluster growth", 
\textit{Surf. Sci.}$\;$ {\bf 604}, L31 (2010).


\bibitem{DJ} K. Donner and P. Jakob, 
``Structural properties and site specific interactions of Pt with the graphene/Ru(0001) moir\'e overlayer", 
\textit{J. Chem. Phys.}$\;$ {\bf 131}, 164701 (2009).


\bibitem{LXYZJFDG} L. W. Liu, W. D. Xiao, K. Yang, L. Z. Zhang, Y. H. Jiang, X. M. Fei, S. X. Du, and H.-J. Gao, 
``Growth and Structural Properties of Pb Islands on Epitaxial Graphene on Ru(0001)", 
\textit{J. Phys. Chem.}$\;$ {\bf 117}, 22652 (2013).

\bibitem{FKYWEHL} Y. Fukamori, M. K\"{o}nig, B. Yoon, B. Wang, F. Esch, U. Heiz, and U. Landman, 
``Fundamental Insight into the Substrate-Dependent Ripening of Monodisperse Clusters", 
\textit{ChemCatChem}$\;$ {\bf 5}, 3330 (2013).

\bibitem{STHSX} L. Semidey-Flecha, D.Teng, B.F. Habenicht, D.S. Sholl, and Y. Xu, 
``Adsorption and diffusion of the Rh and Au adatom on graphene moir\'e/Ru(0001)", 
\textit{J. Chem. Phys.}$\;$ {\bf 138}, 184710 (2013).

\bibitem{PLRMBK} M. Petrovic, P. Lazic, S. Runte, T. Michely, C. Busse, and M. Kralj, 
``Moir\'e-regulated self-assembly of cesium adatoms on epitaxial graphene", 
\textit{Phys. Rev. B}$\;$ {\bf 96}, 085428 (2017).

\bibitem{YKLKLP} J. Yang, K. Kim, Y. Lee, K. Kim, W.C. Lee, and J. Park, 
``Self-organized growth and self-assembly of nanostructures on 2D materials", 
\textit{FlatChem}$\;$ {\bf 5}, 50 (2017).

\bibitem{LYZYBGL} J. Lu, P.S.E. Yeo, Y. Zheng, Z. Yang, Q. Bao, C.K. Gan, and K.P. Loh, 
``Using the graphene Moir\'e pattern for the trapping of C60 and homoepitaxy of graphene", 
\textit{ACS Nano}$\;$ {\bf 6}, 944 (2012).

\bibitem{DXS} S. Dai, Y. Xiang, and D.J. Srolovitz,
``Twisted bilayer graphene: Moir\'e with a twist", 
\textit{Nano Letters}$\;$ {\bf 16}, 5923 (2016).

\bibitem{JJB} S.K. Jain, V. Juricic, and G.T. Barkema,
``Structure of twisted and buckled bilayer graphene", 
\textit{2D Materials}$\;$ {\bf 4}, 015018 (2017).

\bibitem{WB} J. Wintterlin and M.L. Bocquet,  
``Graphene on metal surfaces", 
\textit{Surf. Sci.}$\;$ {\bf 603}, 1841 (2009).

\bibitem{MSPOM} P. Merino, M. Svec, A.L. Pinardi, G. Otero, and J.A. Martin-Gago,  
``Strain-driven Moir\'e superstructures of epitaxial graphene on transition metal surfaces", 
\textit{ACS Nano}$\;$ {\bf 5}, 5627 (2011).

\bibitem{EGW} M.I. Espanol, D. Golovaty, and J.P. Wilber,  
``Discrete-to-continuum modelling of weakly interacting incommensurate two-dimensional lattices", 
\textit{Proc. Royal Soc. A}$\;$ {\bf 474}, 0612 (2018).

\bibitem{VK} D.G. Vlachos and M.K. Katsoulakis, ``Derivation and validation of mesoscopic theories for diffusion of interacting molecules",
\textit{Phys. Rev. Lett.}$\;$ {\bf 85}, 3898 (2000).

\bibitem{LBVK} R. Lam, T. Basak, D. G. Vlachos, and M. A. Katsoulakis, ``Validation of mesoscopic theory and its application to computing concentration
dependent diffusivities", \textit{J. Chem. Phys.}$\;$ {\bf 115}, 11278 (2001).

\bibitem{CV} A. Chatterjee and D.G. Vlachos, ``Continuum mesoscopic framework for multiple interacting species and processes on multiple site types and/or
crystallographic planes", \textit{J. Chem. Phys.}$\;$ {\bf 127}, 034705 (2007).

\bibitem{CTT} M.G. Clerc, E. Tirapegui, and M. Trejo, ``Pattern formation and localized structures in monoatomic layer deposition", 
\textit{Eur. Phys. J. Special Topics}$\;$ {\bf 146}, 407 (2007).

\bibitem{W} D. Walgraef, ``Reaction-diffusion approach to nanostructure formation and texture evolution in adsorbed monoatomic layers", 
\textit{Intl. J. Quant. Chem.}$\;$ {\bf 98}, 248 (2004).

\bibitem{SuoLu} Z. Suo and W. Lu, ``Forces that Drive Nanoscale Self-Assembly on Solid Surfaces", \textit{J. Nanoparticle Res.}$\;$, {\bf 2}, 333 (2000).

\bibitem{LuKim} W. Lu and D. Kim, ``Engineering nanophase self-assembly with elastic field", \textit{Acta Mater.}$\;$ {\bf 53}, 3689 (2005). 

\bibitem{JW} W.C. Johnson and S.M. Wise, ``Phase decomposition of a binary thin film on a patterned substrate", 
\textit{Appl. Phys. Lett.}$\;$ {\bf 81}, 5 (2002).


\bibitem{Painter} K.J. Painter, ``Continuous models for cell migration in tissues and applications to cell sorting via differential chemotaxis", 
\textit{Bull. Math. Biology}$\;$ {\bf 71}, 1117 (2009). 

\bibitem{KK} A.P. Krekhov and L. Kramer, ``Phase separation in the presence of spatially periodic forcing", 
\textit{Phys. Rev. E}$\;$ {\bf 70}, 061801 (2004).


\bibitem{H} K. Hermann, ``Periodic overlayers and moir\'e patterns: theoretical studies of geometric properties", 
\textit{J. Phys.: Condens. Matter}$\;$ {\bf 24}, 314210 (2012).

\bibitem{SMKB} M. Le Ster, T. Maerkl, P.J. Kowalczyk, and S.A. Brown, ``Moir\'e patterns in van der Waals heterostructures", 
\textit{Phys. Rev. B}$\;$ {\bf 99}, 075422 (2019).

\bibitem{MPSO} M. Szendro, A. Palinkas, P. Sule, and Z. Osvath, ``Anisotropic strain effects in small-twist-angle graphene on graphite", 
\textit{Phys. Rev. B}$\;$ {\bf 100}, 125404 (2019).


\bibitem{CWCF} D.A. Cosma, J.R. Wallbank, V. Cheianov, and V.I. Fal'ko, ``Moir\'e pattern as a magnifying glass for strain 
and dislocations in van der Waals heterostructures", \textit{Faraday Discussions}$\;$ {\bf 173}, 137 (2014).

\bibitem{WMCF} J.F. Wallbank, M. Mucha-Kruczynski, X. Chen, and V.I. Fal'ko, ``Moir\'e superlattice effects in graphene/boronnitride
van der Waals heterostructures", \textit{Annalen der Physik}$\;$ {\bf 527}, 359-376 (2015).

\bibitem{SuttonChen} A.P. Sutton and J. Chen, ``Long-range Finnis-Sinclair potentials", \textit{Phil. Mag. Lett.}$\;$ {\bf 61}, 139 (1990).


\bibitem{RM} H. Rafii-Tabar and G.A. Mansoori,  ``Interatomic potential models for nanostructures", 
\textit{Encyclopedia of nanoscience and nanotechnology}$\;$ {\bf 4}, 231 (2004).

\bibitem{LC} S.Y. Liem and K.-Y. Chan, ``Effective pairwise potential for simulations of adsorbed platinum", \textit{Molecular Phys.}$\;$ {\bf 86}, 
939 (1995). 

\bibitem{HVFN} H. Heinz, R.A. Vaia, B.L. Farmer, and R.R. Naik, ``Accurate simulation of surfaces and interfaces of face-centered cubic metals using 12-6
and 9-6 Lennard-Jones potentials", \textit{J. Phys. Chem. C}$\;$ {\bf 112}, 17281 (2008).

\bibitem{AV} N.M. Abukhdeir and D.G. Vlachos, ``Nanoscale surface pattern evolution in heteroepitaxial bimetallic films", 
\textit{ACS Nano}$\;$ {\bf 5}, 7168 (2011).

\bibitem{honey} See, for instance, \url{https://www.weizmann.ac.il/condmat/oreg/sites/condmat.oreg/files/uploads/2015/tutorial_1.pdf}.

\bibitem{BRHB} A. Barreiro, R. Rurali, E.R. Hernandez, and A. Bachtold, ``Structured graphene devices for mass transport", 
\textit{Small}$\;$ {\bf 7}, 775 (2011).

\bibitem{SV} D. Solenov and K.A. Velizhanin, ``Adsorbate transport on graphene by electromigration", \textit{Phys. Rev. Lett.}$\;$ {\bf 109}, 095504 (2012).

\bibitem{MySurfSci} M. Khenner, ``Electromigration-guided composition patterns in thin alloy films: a computational
study", \textit{Surf. Sci.}$\;$ {\bf 698}, 121611 (2020).

\bibitem{MyPRMat} M. Khenner, ``Directed long-range transport of a nearly pure component atom clusters by the electromigration of a binary 
surface alloy", \textit{Phys. Rev. Materials}$\;$ {\bf 5}, 024001 (2021).








\end{thebibliography}
\end{document}